\documentstyle[twoside,fleqn,epsfig]{article}

\def\be{\begin{equation}}     
\def\ee{\end{equation}}
\def\bea{\begin{eqnarray}}
\def\eea{\end{eqnarray}}
\newcommand{\lsim}{\mathrel{\mathop{\kern 0pt \rlap
  {\raise.2ex\hbox{$<$}}}
  \lower.9ex\hbox{\kern-.190em $\sim$}}}
\newcommand{\gsim}{\mathrel{\mathop{\kern 0pt \rlap
  {\raise.2ex\hbox{$>$}}}
  \lower.9ex\hbox{\kern-.190em $\sim$}}}

\newcommand{\AmS}{{\protect\the\textfont2
  A\kern-.1667em\lower.5ex\hbox{M}\kern-.125emS}}
\normalsize
\begin{document}
\baselineskip=0.65cm

\vspace{0.5cm}
\begin{center}  
\large
{\bf New search for correlated $e^+e^-$ pairs in the $\alpha$ decay of $^{241}$Am} \\

\large

\vspace{0.5cm}

R.~Bernabei$^{1,2}$,
P.~Belli$^{1,2}$,
F.~Cappella$^{3,4}$,
V. Caracciolo$^{5}$,
S. Castellano$^{5}$, 
R.~Cerulli$^{5}$,
C.J. Dai$^{6}$,
A.~d'Angelo$^{3,4}$,
A.~Di Marco$^{1,2}$,
H.L. He$^{6}$,
A.~Incicchitti$^{3,4}$,
M.~Laubenstein$^{5}$,
X.H. Ma$^{6}$,
F. Montecchia$^{1,7}$, 
X.D. Sheng$^{6}$,
V.I.~Tretyak$^{8}$,
R.G. Wang$^{6}$,
Z.P. Ye$^{6,9}$

\normalsize
\vspace{0.5cm}
$^1${\it INFN sezione Roma ``Tor Vergata'', I-00133 Rome, Italy}\\
\vspace{1mm}
$^2${\it Dipartimento di Fisica, Universit\`a di Roma ``Tor Vergata'', I-00133 Rome, Italy}\\
\vspace{1mm}
$^3${\it INFN sezione Roma, I-00185 Rome, Italy}\\
\vspace{1mm}
$^4${\it Dipartimento di Fisica, Universit\`a di Roma ``La Sapienza'', I-00185 Rome, Italy}\\
\vspace{1mm}
$^5${\it INFN, Laboratori Nazionali del Gran Sasso, I-67100 Assergi (AQ), Italy}\\
\vspace{1mm}
$^6${\it IHEP, Chinese Academy, P.O. Box 918/3, Beijing 100039, China}\\
\vspace{1mm}
$^7${\it Dipartimento di Ingegneria Civile e Ingegneria Informatica, Universit\`a di Roma ``Tor Vergata'', Italy}\\
\vspace{1mm}
$^8${\it Institute for Nuclear Research, MSP 03680 Kyiv, Ukraine}\\
\vspace{1mm}
$^9${\it University of Jing Gangshan, Jiangxi, China}\\
\vspace{1mm}
\end{center}

\normalsize

\vspace{0.3cm}
\begin{abstract} 
A new search for production of correlated $e^+e^-$ pairs in the $\alpha$ decay of 
$^{241}$Am has been carried out deep underground at the Gran Sasso 
National Laboratory of the I.N.F.N. by using pairs of NaI(Tl) detectors of the DAMA/LIBRA set-up.
The experimental data show an excess of double coincidences of events with energy around 511 keV in faced pairs of detectors,
which are not explained by known side reactions. This measured excess gives a relative activity 
$\lambda = (4.70 \pm 0.63) \times 10^{-9}$ for the Internal Pair Production (IPP)
with respect to alpha decay of $^{241}$Am; this value is of the same order of magnitude as previous determinations. 
In a conservative approach the upper limit $\lambda < 5.5 \times 10^{-9}$ (90\% C.L.) can be 
derived. It is worth noting that this is the first result on IPP obtained in an underground
experiment, and
that the $\lambda$ value obtained in the present work is independent on the
live-time estimate.
\end{abstract}

\normalsize

\vspace{0.3cm}
\noindent {\it PACS}: {23.20.Ra, Internal pair production --} 
      {27.90.+b, $A\ge220$ --}   
      {23.60.+e, Alpha decay --} 
      {29.40.Mc, Scintillation detectors}

\vskip 0.4cm

\section{Introduction} 
\label{par1}

In the seventies experimental and theoretical investigations of weak interactions in nuclei suggested the presence of various electromagnetic phenomena, 
which modify the basic decay processes. In the conventional perturbation treatment of the weak interactions these phenomena are attributed to higher-order 
terms. These higher-order terms, accompanying $\beta$ decay and shell-electron capture, are \cite{Pet66,Lju73}: (i) internal bremsstrahlung; (ii) 
ionization and excitation of the electron cloud; (iii) internal pair production (IPP). These higher-order effects are much less intense than the 
first-order processes; for example, in the case of the internal bremsstrahlung about one photon is produced per 100 $\beta$ decays, and the internal 
ionization process only occurs about once per 10$^4$ decays. 
The first estimate of the contribution of the IPP process in the $\beta$ decay was performed by Arley and Moller \cite{Arl38} and, independently, by Tisza 
\cite{Tis37}; these studies were refined in the following years, and experimentally tested by Greenberg and Deutsch \cite{Gre56}. 

In 1973 \cite{Lju73} and in the following years, the IPP process was also investigated in 
the $\alpha$ decay, where this effect could give additional information 
on the higher-order phenomena accompanying nuclear transformations. In Ref. \cite{Lju73}, a theoretical estimate of the $e^+e^-$ production due to 
bremsstrahlung $\gamma$ emitted by $\alpha$ particles in the $^{241}$Am decay was given: $\lambda = \frac{A_{e^+e^-}}{A_{\alpha}}=1.2 \times 10^{-9}$, 
where $A_{e^+e^-}$ and $A_{\alpha}$ are the $e^+e^-$ and $\alpha$ activities, respectively. This estimate was in reasonable agreement with the 
experimental result, obtained in the same work by investigating double coincidences of the two 511 keV $\gamma$ quanta 
produced by the positron annihilation; the set-up did not allow the detection of the emitted $e^+e^-$ pairs. 
In particular, a source of $^{241}$AmO$_2$ (deposited on top of Pt backing and covered with thin Au layer) was mounted between Ge(Li) 40 cm$^3$ 
and NaI(Tl) 80 cm$^3$ detectors. The mechanical support of the source was made of a stainless 
steel backing (without direct access of $\alpha$'s to the stainless steel); Pb and V foils were used to stop the emitted $\alpha$ particles. 
The observation of two 511 keV $\gamma$ quanta in coincidence between the Ge(Li) and NaI(Tl) 
gave for the probability of the $e^+e^-$ emission the value: $\lambda = (3.1 \pm 0.6) \times 10^{-9}$. 
The authors excluded (or took into account) various background processes which could 
imitate the observation, such as: i) $e^+e^-$ pair creation by cosmic rays and high energy $\gamma$'s from natural radioactivity; 
ii) high energy external bremsstrahlung from $\alpha$'s; iii) ($\alpha$,n), ($\alpha$,p) reactions on Pt, Pb, V, O; iv)
excitation of high energy nuclear levels in ($\alpha$,$\alpha$') scattering; v) presence of $^{65}$Zn and other long lived $\beta^+$ emitters;
vi) decay of $^{241}$Am to high energy levels of $^{237}$Np.

In 1978 the theoretical value of the
probability of the process in $^{241}$Am was re-estimated to be \cite{Pis78}: $2.3 \times 10^{-9}$.
Further theoretical studies have also been carried out; in fact, while the theory of the IPP in the 
$\beta$ decay seemed to be clear, in the $\alpha$ decay it  
required more attention \cite{Lju73,Pis78,Pre79,Chu81,Log82}. Essentially two modes have been suggested for the IPP process in $\alpha$ decay: (i) the $\alpha$ 
particles are accelerated in the Coulomb field of the daughter nucleus and emit bremsstrahlung $\gamma$'s, with energy above $2m_ec^2$, giving rise to $e^+e^-$ pairs; 
(ii) the daughter nucleus originates in an excited state, and the resulting $\gamma$ quantum (real or virtual) is converted to a $e^+e^-$ 
pair. Both processes are possible, as pointed out in Ref. \cite{Log82}; the process (i) proposed in Refs. \cite{Lju73,Pis78} gives a theoretical 
estimate\footnote{For completeness we recall that in \cite{Pre79,Chu81} 
the $\lambda$ value for this process was estimated to be $\sim 10^{-20}$, but \cite{Log82} has shown that this value was incorrect because it didn't take 
into account the full
transition energy of the $\alpha$ particles.}  
$\lambda \approx 10^{-9}$ (as forementioned), whereas the process (ii) discussed in Refs. \cite{Pre79,Chu81} gives $\lambda \approx 10^{-12}$. Therefore, 
according to Ref. \cite{Log82} the dominant contribution to the IPP in the 
$\alpha$ decay of $^{241}$Am arises from the mode (i). 

In 1986 a new dedicated experiment \cite{Sta86} was performed in order to investigate the 
result of Ref. \cite{Lju73}. 
In this experiment, three $\alpha$ emitters were studied: $^{210}$Po, $^{239}$Pu, $^{241}$Am. 
Pb and Cu absorbers were used to stop $\alpha$ and $\beta^+$ particles; the 511 keV double coincidences between 
a Ge(Li) 70 cm$^3$ and a NaI(Tl) 785 cm$^3$ detectors, placed in a low background shield, were observed. For all the sources, 
the process was observed with probabilities: $\lambda(^{210}$Po$) = (5.3 \pm 1.7) \times 10^{-9}$, $\lambda(^{239}$Pu$) = (7 \pm 9) \times 10^{-9}$, 
$\lambda(^{241}$Am$) = 
(2.15 \pm 0.25) \times 10^{-9}$, respectively. Pollution of the $^{241}$Am source by $^{152}$Eu and $^{154}$Eu, and pollution of 
the $^{239}$Pu by $^{241}$Am, $^{234m}$Pa, $^{214}$Bi, $^{208}$Tl, $^{137}$Cs were measured, while in the $^{210}$Po source radioactive impurities were not identified. 
Moreover, any $\alpha$ decay of $^{241}$Am 
to the high energy ($ > 1022$ keV) excited states of $^{237}$Np 
was excluded (intensity $ < 10^{-9}$).
The obtained results were in agreement with the refined theoretical expectations of Ref. \cite{Pis78} (see Table \ref{tab1}), 
and in particular the previous positive result \cite{Lju73} on $^{241}$Am was confirmed.

\begin{table*}[!ht]
\caption{Experimental results and theoretical estimates of
$\lambda$ for the IPP process in $\alpha$ decays. The superscript $^a$ identifies the 
$\lambda$ value derived from Ref. \cite{Asa90} when adopting the procedure
described in the text.}
\begin{center}
\resizebox{\textwidth}{!}{\begin{tabular}{l|cccc|ccc}
\hline \hline 
Source & \multicolumn{4}{c|}{Experiment} & \multicolumn{3}{c}{Theory} \\
       &  $\lambda$ ($\times 10^{-9}$)  & Detectors & Year & Ref. & $\lambda$ ($\times 10^{-9}$) & Year & Ref. \\
\hline \hline
 & & & & & & & \\
$^{210}$Po & $5.3 \pm 1.7$   & NaI(Tl)+Ge(Li) & 1986 & \cite{Sta86} & 4.4 & 1978 & \cite{Pis78} \\ 
 & & & & & & & \\
\hline
 & & & & & & & \\
$^{239}$Pu & $7 \pm 9$       & NaI(Tl)+Ge(Li) & 1986 & \cite{Sta86} & 2.2 & 1978 & \cite{Pis78} \\ 
 & & & & & & & \\
\hline
 & & & & & & & \\
$^{241}$Am & $3.1 \pm 0.6$   & NaI(Tl)+Ge(Li) & 1973 & \cite{Lju73} & 1.2 & 1973 & \cite{Lju73} \\ 
           & $2.15\pm0.25$   & NaI(Tl)+Ge(Li) & 1986 & \cite{Sta86} & 2.3 & 1978 & \cite{Pis78} \\ 
           & $1.8 \pm 0.7^a$ & Plastics+Ge    & 1990 & \cite{Asa90} &     &      &              
\\ 
           & $4.70 \pm 0.63$ & NaI(Tl) pairs       & 2013 & this work    &     &      &              
\\ 
 & & & & & & & \\
\hline \hline
\end{tabular}}
\label{tab1}
\end{center}
\end{table*}

In the late eighties, some anomalous phenomena were observed in collision of heavy ions at GSI and in collision of heavy ions with $e^+$ \cite{Ser87},
and it was suggested the possible existence of an hypothetical neutral $X$ particle, whose decay in 
$e^+e^-$ pair may give rise to the observed peaks in the measured $e^+$ and $e^-$ spectra. 
In particular, in Ref. \cite{Ino90} it was proposed a new model where the pair production would be due to
a QED strong 
coupling phase: heavy nuclei have a surface region with a strong static electric field that can be perturbed by the 
passage of charged particles (e.g. $\beta$ 
and $\alpha$). The perturbation causes a fluctuation in the electric field, that would produce a strong non-static 
electromagnetic field which may be source of strong 
coupling vacuum; the vacuum bubbles can decay in positronium states, which are $e^+e^-$ bound 
states. In Ref. \cite{Ino90} some experimental tests were suggested for this model; 
one of them is the search for double coincidence events produced by positron annihilation. Thus, in this scenario the production of positronium states in $\alpha$ 
decay of heavy nuclei may be a concurrent process with respect to the IPP one. 
However, since the IPP process is better stated, in the following we will mainly refer the results to it.

In 1990 an experiment
was performed \cite{Asa90} with the aim to test the model of Ref. \cite{Ino90}. 
Two thin plastic scintillators registered $e^+e^-$ pairs emitted from an $^{241}$Am source, and two Ge detectors of 198 cm$^3$ each registered two 511 keV 
$\gamma$ quanta. The $^{241}$Am source was sandwiched by a pair 
of 10 $\mu$m Ti foils followed by two aluminized mylar foils of 30 $\mu$m; two plastic scintillators were placed after them. 
During 541.7 h live time, the coicidences between 
$\simeq$ 511 keV events in the two Ge detectors were 
466, most of them were explained as due to background from $\beta^+$ decay of $^{26m}$Al produced by 
$^{23}$Na($\alpha$,n)$^{26m}$Al from Sodium contamination;
the presence of Sodium in the vicinity of $^{241}$Am source was also confirmed by 
the observation of 
the $\gamma$ lines at 1809 and 1130 keV emitted by the $^{26}$Mg$^*$ produced in the process: 
$^{23}$Na($\alpha$,p)$^{26}$Mg$^*$.
When requiring energy releases in two plastics: $E_1$ and $E_2$ greater than  30 keV,
just 6 coincidences survive. 
Finally, only one event remains when  
the last requirement $E_1 \approx E_2$ -- as expected
by the model of Ref. \cite{Ino90} -- is adopted; hence, an upper limit on the probability of the
production of a hypothetical neutral particle with mass above
1.4 MeV/c$^2$ decaying in $e^+e^-$, was set: $<1.5 \times 10^{-9}$ (95\% confidence level, 
C.L.).
The previous 
experimental results regarding the IPP process \cite{Lju73,Sta86} (and also the theoretical estimates) were missed in this work.
However, in case the IPP process is instead considered, the requirement $E_1 \approx E_2$ should be released, and thus 
the useful number of 511 keV double coincidences from Ref. \cite{Asa90} 
is 6 events; considering this fact, one can derive    
a $\lambda$ value: $(1.8 \pm 0.7) \times 10^{-9}$, compatible with previous observations.
It should be, however, noted that -- since the efficiency in \cite{Sta86} was calculated by simulating 
the $X$ decay with mass of $m_X >$ 1.4 MeV -- the derived value of $\lambda$ has to be considered approximate.

Thus, considering the present situation (see Table \ref{tab1}), a new search for $e^+e^-$ pairs in the $^{241}$Am $\alpha$ decay has 
been carried out deep underground. 
The obtained results are described in the following.

\section{The experimental set-up and the data taking conditions}

The results presented in the following have been obtained through dedicated measurements with some pairs of the NaI(Tl) 
detectors of the DAMA/LIBRA set-up \cite{perflibra,mod1,mod2,papep,cnc,NewPMTs}, whose description, radiopurity and main 
features are discussed in details in Refs. \cite{perflibra,mod2,NewPMTs}. We just recall that the sensitive 
part of this set-up is made of 25 highly radiopure NaI(Tl) crystal scintillators 
(5-rows by 5-columns matrix); each NaI(Tl) detector has 9.70 kg mass and a size of ($10.2 \times 10.2 \times 25.4$) cm$^{3}$. The bare crystals are enveloped in 
Tetratec-teflon foils and encapsulated in radiopure OFHC Cu housings. In each detector two 10 cm long special quartz light guides also act as 
optical windows on the 
two end faces of the crystal and are coupled to two low background photomultipliers working in coincidence at single photoelectron threshold. The data taking 
considered here has been performed in the new DAMA/LIBRA configuration after the upgrading occurred in fall 2010 when all the PMTs were replaced by new ones with higher 
quantum efficiency, specially developed by HAMAMATSU; details can be found in Ref. \cite {NewPMTs}. The light response in this experimental configuration is typically 
6--10 ph.e./keV depending on the detector. The detectors are housed in a sealed low radioactivity copper box installed in the center 
of a low radioactivity
Cu/Pb/Cd-foils/polyethylene/paraffin shield; moreover, about 1 m concrete (made from the Gran Sasso rock material) almost fully 
surrounds (mostly outside the barrack) 
this passive shield, acting as a further neutron moderator. The copper box is maintained in HP Nitrogen atmosphere in slightly overpressure with respect to the 
external environment; it is part of the threefold-level sealing system which excludes the detectors from environmental air. For more details see e.g. 
Refs. \cite{perflibra,mod2,NewPMTs}.

In the particular measurements dedicated to the present study an $^{241}$Am source was placed in the middle of each used NaI(Tl) pair for a total of 6 sources used.
The sources have been produced by CEA-DAMRI (nowadays called CERCA-LEA) and are identified as AM241-EGSA15. These sources are made of a 
thin circular layer of active AmO$_2$ ($\phi \sim$ 3 mm) 
heat sealed by two thin mylar foils. They have been stored deep underground for $\gsim$ 10 yr; their present mean activity is $\sim$ 
35 kBq. 

The detailed decay scheme of $^{241}$Am can be found in Ref. \cite{Bas06}.
The energy spectrum measured by the DAMA/LI\-BRA detectors, when irradiated by the $^{241}$Am sources, has been accurately 
reproduced by a suitable MonteCarlo simulation, which takes into account the decay scheme of $^{241}$Am \cite{Bas06}
and the activities of the sources. It is worth noting that about 99.91\% of the $^{241}$Am total rate
is due to events below 90 keV. 
Thus, in order to reduce the acquisition dead-time in our measurements the energy threshold for each 
detector has been set at $\approx$90 keV 
to exclude most of the events from the $^{241}$Am which are out of the region of interest for the present investigation. 
The accuracy of the MonteCarlo simulation allows us to be confident with the standard procedures used for the determination
of the data acquisition live time.
The 
electronic chain was modified with respect to the usual one \cite{perflibra} by excluding the 500 $\mu$s blocking time and by using either twelve 
detectors (six pairs) in the first
running period or six detectors (three pairs) in the second one. 
An improvement of a factor $\approx 2$ has been achieved as regards the ratio of the live time over elapsed one 
in the second dedicated run, where just three pairs have been used (those with the lower number of coincidences in the first run). 

In both running periods each pair has the detectors of the other used pairs as anti-coincidences; 
this implies two different coincidence backgrounds for a same pair in the two different 
running periods (see  
Tables \ref{tab2} and \ref{tab3}).
Finally, the energies of the acquired events  
were read out from the ADCs (see e.g. Ref. \cite{perflibra}).

\section{Experimental results}

The experimental procedure is based on the detection of the two $\gamma$'s from the positron annihilation 
through the study of double coincidences in pairs of faced highly radiopure NaI(Tl) detectors.

\subsection{The first dedicated run}

As mentioned above, for practical reasons in this data taking six pairs of faced highly radiopure NaI(Tl) 
detectors (i.e. 12 of the 25 available) of the DAMA/LIBRA set-up have been used. The activity, $A$, of each used 
$^{241}$Am source is given in Table \ref{tab2}; the total live 
time was 1.29 d. 

In the data analysis an energy window (465--557) keV -- practically $\pm 2\sigma$ around 511 keV -- 
has been considered for the coincidences for both the detectors of a pair.
Since no $\alpha$ decay of $^{241}$Am to an excited level of $^{237}$Np with energy larger than 0.8 MeV 
has been observed, to our knowledge and in the decay scheme of $^{241}$Am \cite{Bas06}\footnote{One should note 
that in the list of $^{241}$Am $\gamma$ quanta \cite{TOI98} there are unplaced $\gamma$'s with low intensity and higher 
energies; however, even the most energetic one, having $E_\gamma=1014.5$ keV (intensity $6.4\times10^{-10}$), 
has not enough energy to create $e^+e^-$ pair.},
a contribution of $^{241}$Am decay in the applied coincidence energy windows is not expected.
Hence, a possible excess with respect to the data obtained when running without sources
is due to the sources but not directly to the known $^{241}$Am decay scheme \cite{Bas06}.

The obtained results are summarized in Table \ref{tab2}, where
\begin{table}[!ht]
\caption{Experimental results of the first running period: i)
$D1-D2$ identifies the NaI(Tl) detectors in each pair according to the numbering given in Ref. \cite{mod1}; 
ii) $A$ is the activity of the used $^{241}$Am 
source at time of measurements (error given by the producer: 3.5\%); iii) $n_1$ is the number of the double coincidences measured in the (465--557) 
keV energy window during 1.29 d 
of live time in presence of the $^{241}$Am sources; iv) $n_2$ is the 
number of double coincidences measured in the same conditions during 
24.6 d live time without sources; v) $s$ is the excess rate of double coincidences 
derived according to the statistical procedure given in the 
Appendix. The upper limits are at 90\% C.L. See text.}
\begin{center}
\begin{tabular}{ccccc}
\hline
\hline 
  $D1-D2$ &  $A$    &  $n_1$         & $n_2$           & $s$     \\
          & (kBq)   & ($T_1$=1.29 d) & ($T_2$=24.6 d)  & (counts/day)   \\
\hline \hline
  1-2   & 34.8  &  5             &  23             &  2.9$^{+2.0}_{-1.5}$ \\
  6-7   & 40.8  & 14             &  26             &  9.8$^{+3.2}_{-2.7}$ \\
  9-10  & 31.7  & 10             &  25             &  6.7$^{+2.7}_{-2.2}$ \\
  11-12 & 34.6  &  1             &  28             &  $<2.4$              \\
  16-17 & 29.5  &  5             &  22             &  3.0$^{+2.0}_{-1.5}$ \\
  19-20 & 29.4  & 10             &  17             &  7.0$^{+2.7}_{-2.2}$ \\
 \hline
Total   & 200.8 & 45             & 141             & 29.2$\pm5.2$         \\
\hline
\end{tabular}
\label{tab2}
\end{center}
\end{table}
the number of the double coincidences ($n_1$) measured in the (465--557) keV energy window during 1.29 d of live time ($T_1$) in presence of the $^{241}$Am sources are 
compared with those ($n_2$) recorded in the same energy window during 24.6 d live time ($T_2$) without any source. 

Either the estimate of the coincidence excess rate in the considered energy window, $s$, 
or the limit (90\% C.L.) value have been calculated according to 
the statistical procedure described in the Appendix. 
The obtained $s$ values are compatible; therefore, a cumulative analysis has been performed (last row of Table 
\ref{tab2}). The cumulative data give the result: $s = (29.2 \pm 5.2) $ counts/day, that is $(4.9 \pm 0.9) $ counts/day/pair.

\subsection{The second dedicated run}

The second dedicated run has been performed using the three detectors pairs/sources which had, in the first dedicated run, 
the lowest values of $s$; these pairs are: 1-2, 11-12 and 16-17 (see Table \ref{tab2}). 
The experimental live time in this second dedicated run was 2.63 d. 

The results are summarized in Table \ref{tab3}.
\begin{table}[!ht]
\caption{Experimental results of the second running period. See the caption of Table 1 for a description of the given 
quantities. Here the data taking live time with $^{241}$Am sources is 2.63 d ($T_1$), and without them is 24.6 d ($T_2$). See text.}
\begin{center}
\begin{tabular}{cccccc}
\hline
\hline 
$D1-D2$   & $A$     &  $n_1$         & $n_2$           & $s$             \\
          & (kBq)   & ($T_1$=2.63 d) & ($T_2$=24.6 d)  & (counts/day)    \\
 \\ \hline \hline
  1-2   & 34.8  & 19             & 29              & 6.0$^{+1.8}_{-1.6}$ \\
  11-12 & 34.6  & 15             & 33              & 4.3$^{+1.6}_{-1.4}$ \\
  16-17 & 29.5  & 17             & 29              & 5.3$^{+1.7}_{-1.5}$ \\
 \hline
Total   & 98.9 & 51             & 91              & 15.7$\pm2.7$        \\
\hline
\end{tabular}
\label{tab3}
\end{center}
\end{table}
Also in this case, the obtained $s$ values are well compatible; thus, the cumulative analysis is performed obtaining: $s = (15.7 \pm 2.7) $ counts/day, 
that is $(5.2 \pm 0.9) $ counts/day/pair.

\subsection{The cumulative results}

The excess rates per pair in the region of interest during the two running periods are compatible. 
Thus, the compatibility and the independence between the first and the second dedicated runs 
allow us to perform a combined analysis of the two data sets.

\begin{figure*}[!ht]
\begin{center}
\resizebox{0.9\textwidth}{!}{\includegraphics{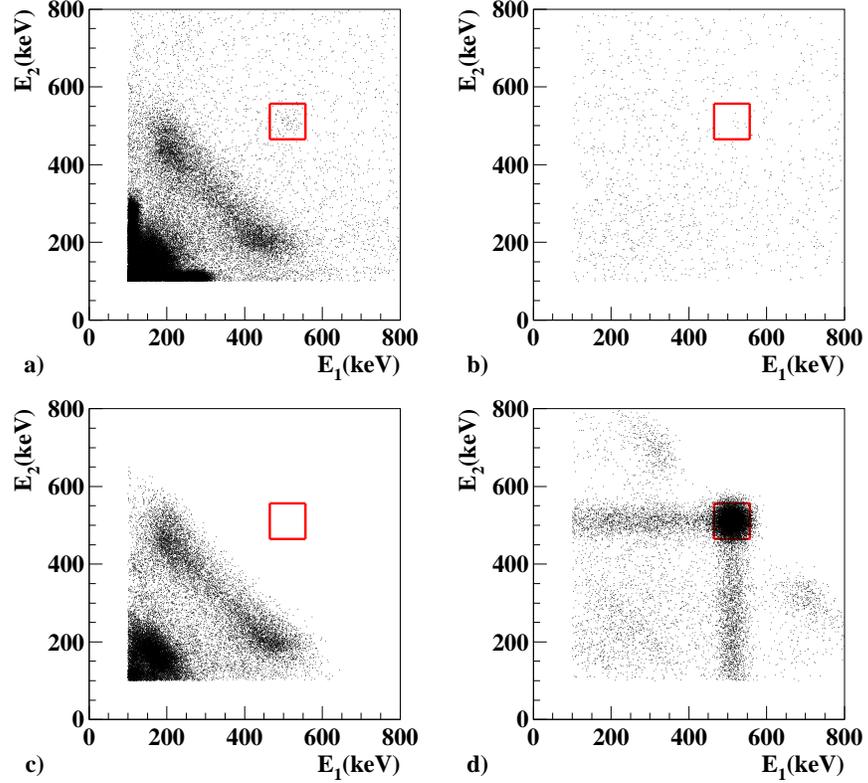}}
\end{center}
\vspace{-0.8cm}
\caption{($Color$ $online$) 
Coincidence scatter plots of the used pairs of faced NaI(Tl) detectors:
a) experimental scatter plot measured during the cumulative 3.92 d live time in presence of the $^{241}$Am sources;
b) experimental scatter plot without sources (the number of displayed events is normalized to the same time as the case a));
c) model scatter plot of the $^{241}$Am decay \cite{Bas06} obtained by simulation (the number of displayed events is normalized to the same time 
   as the case a) and applying the sources activities given in Table \ref{tab2} and \ref{tab3});
d) model scatter plot of the positrons annihilations within the sources obtained by simulation (the number of events in the red box is 100-times those of a)).
The red boxes define the region with $E_1$ and $E_2$ in the energy window (465--557) keV ($\pm 2 \sigma$ around 511 keV);
the spot at 511--511 keV can be recognized in a), while it is absent in b) and c).
See text.}
\label{fig1}
\end{figure*}

In Fig. \ref{fig1}a) the coincidence scatter plot $E_2$ $vs$ $E_1$ of the considered pairs of faced NaI(Tl) detectors
is shown for the cumulative data collected in presence of the $^{241}$Am sources during 3.92 d live time.
The red box defines the region with $E_1$ and $E_2$ in the energy window (465--557) keV ($\pm 2 \sigma$ around 511 keV); the spot 
at 511--511 keV can be recognized.
This excess is not present in the data measured in absence of the sources, Fig. \ref{fig1}b), and in the model of the $^{241}$Am decay 
according to the scheme of Ref. \cite{Bas06}, Fig. \ref{fig1}c).  Most of the other coincidence events outside 
the red box of Fig. \ref{fig1}a) are instead well described by the scatter plots in Fig. \ref{fig1}b) and \ref{fig1}c).
Finally, Fig. \ref{fig1}d) shows the model scatter plot for the case of positrons annihilations within the sources; the 27\% of the 
positrons annihilations give events that are contained in the red box. The scatter plots in Fig. \ref{fig1}b), c) and d) are
suitably normalized (see the respective captions).

The analysis of the double coincidences allows us to point out the presence of $^{243}$Am 
(alpha decaying with half-life 7370 yr) and of $^{154}$Eu (mainly beta decaying with half-life 8.601 yr) isotopes
in the sources. 
Their presence can be justified by side reactions in the production procedures of the $^{241}$Am 
sources. In fact: i) the $^{241}$Am is produced bombarding $^{239}$Pu
nuclides with neutrons; ii) the $^{243}$Am is produced by neutron captures on $^{241}$Am; iii) 
the $^{154}$Eu is produced by induced fissions of $^{239}$Pu.
Moreover, from the study of the experimental cumulative energy spectrum with multiplicity $\ge 1$ shown in the following,
the presence of $^{233}$Pa isotope can be pointed out; it belongs to the $^{241}$Am chain and it is the
only daughter of $^{241}$Am able to produce a not negligible contribution of $\gamma$'s above 100 keV.

\begin{figure*}[!ht]
\begin{center}
\vspace{-0.8cm}
\resizebox{0.45\textwidth}{!}{\includegraphics{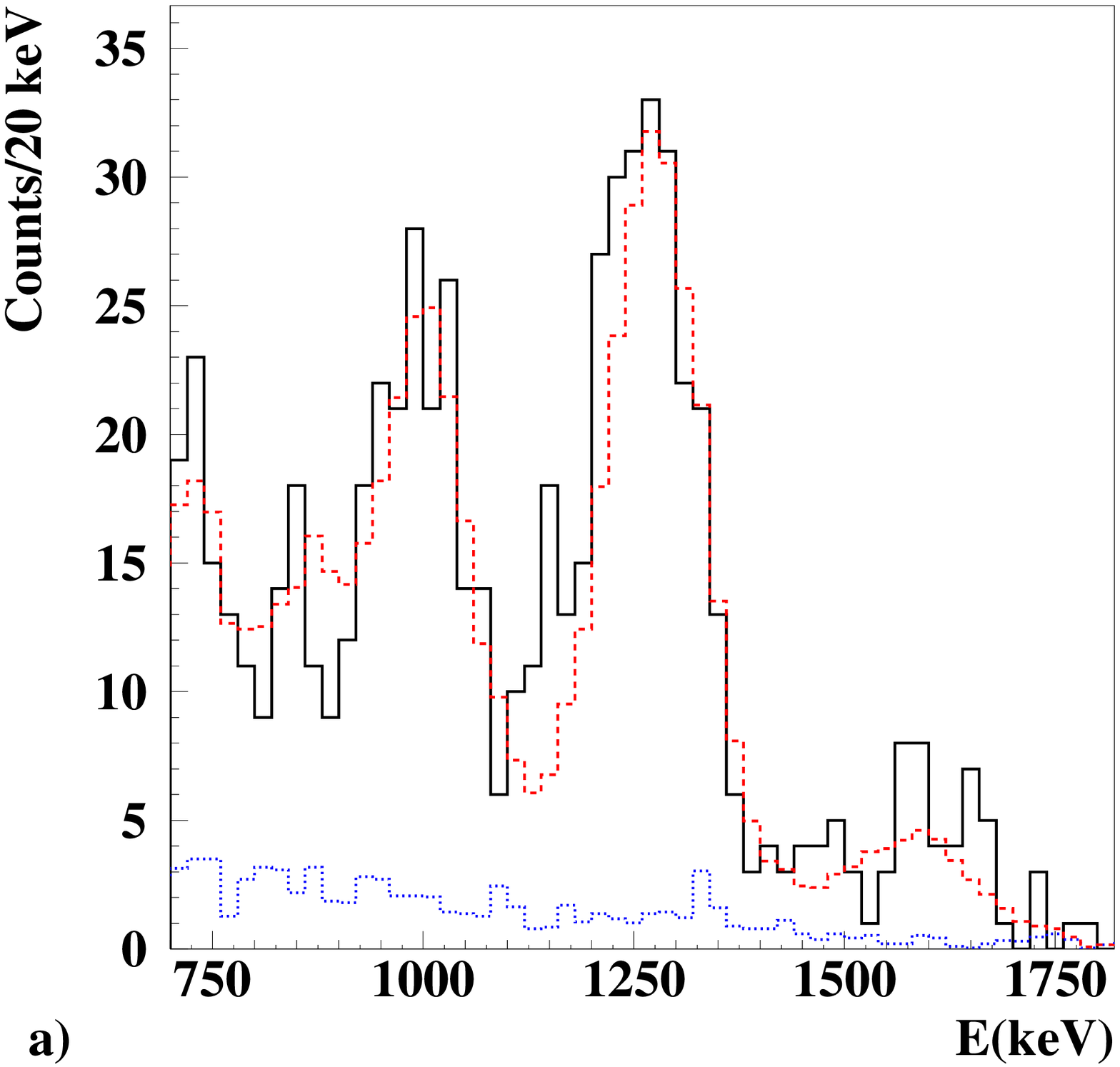}}
\resizebox{0.45\textwidth}{!}{\includegraphics{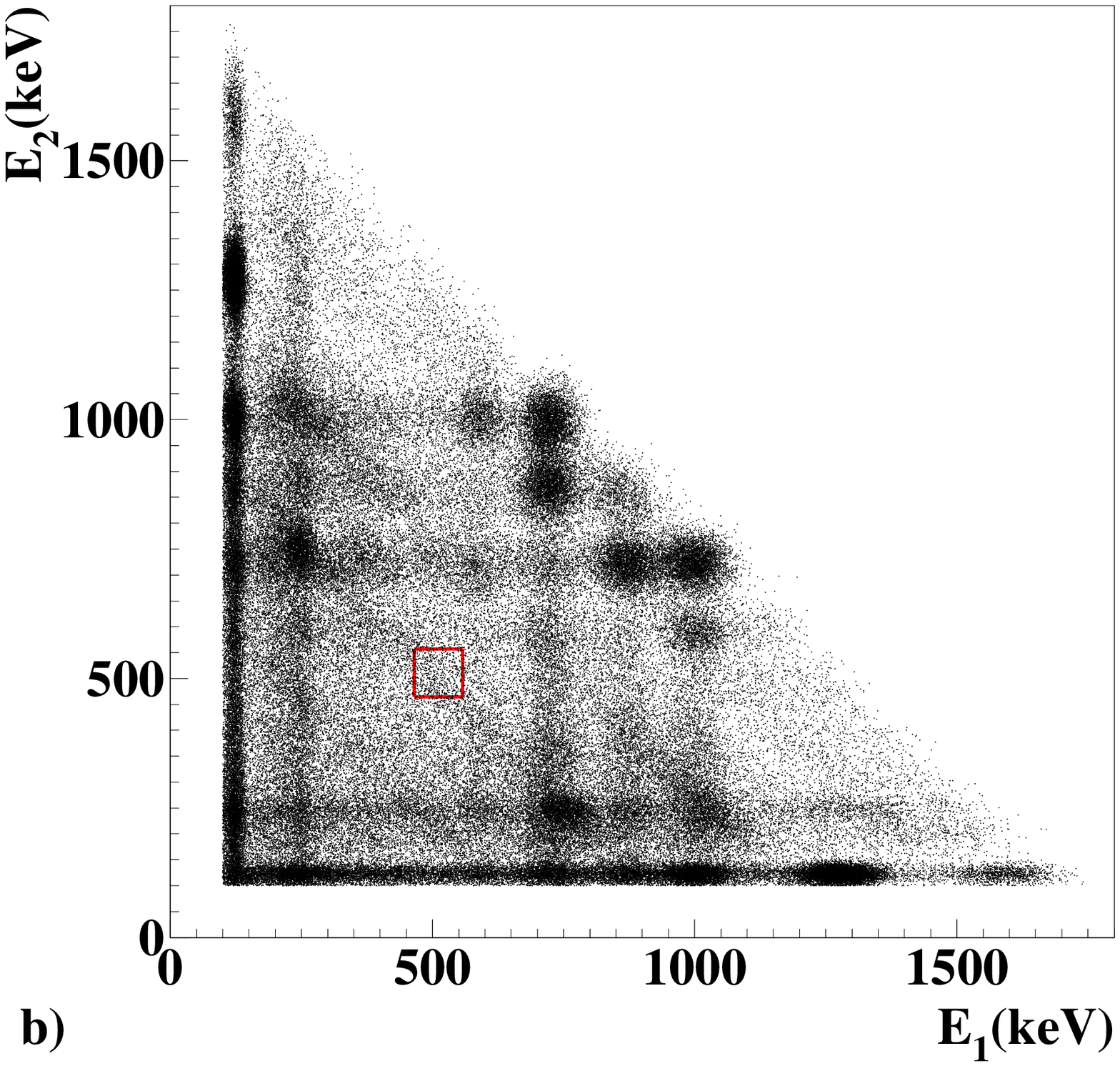}}
\end{center}
\vspace{-0.6cm}
\caption{($Color$ $online$) a) The distribution of $E_2$, requiring that $E_1$ is in the 100--150 keV energy window (see Fig. \ref{fig1}a)),
is shown for the cumulative 3.92 d 
live time in presence of $^{241}$Am sources (solid black histogram) and for 24.6 d 
without the sources, suitably normalized (dotted histogram -- blue online).
The presence of the $^{154}$Eu peaks is clearly evident. The dashed curve (red online) is the result of the best
fit, that includes the contributions of the set-up (dotted histogram -- blue online)
and of the $^{154}$Eu decay model.
The contribution from $^{241}$Am decay is negligible in this energy window.
b) model scatter plot of the $^{154}$Eu decay obtained by the simulation. The number of plotted events is arbitrary.
The red box defines the region with $E_1$ and $E_2$ 
in the energy window (465--557) keV ($\pm 2 \sigma$ around 511 keV).}
\label{fig_eu}
\end{figure*}

The $^{154}$Eu can be mainly identified by the detection of a $\gamma$ of 123 keV in coincidence with $\gamma$'s of
1005 keV (this occurs in the 8.2\% of the $^{154}$Eu decays), 1274 keV (15.8\%), 1596 keV (0.8\%).
Hence, selecting a slice in the double coincidence scatter plot
of Fig. \ref{fig1}a) with $E_1$ in the 100--150 keV energy window, the spectrum of $E_2$ 
is produced and reported in Fig. \ref{fig_eu}a). The best fit\footnote{The contribution 
from $^{241}$Am decay is negligible in this energy window.}, which includes the contributions of the set-up and of the 
$^{154}$Eu decay model (also see Fig. \ref{fig_eu}b)), provides $(1.70\pm0.08) \times 10^4$ $^{154}$Eu decays.
This corresponds to an average activity of $^{154}$Eu equal to $(12.6\pm0.6)$ mBq in each source; 
thus, the expected number of coincidences with $E_1$ and $E_2$ in the energy window (465--557) keV ($\pm 2 \sigma$ around 511 keV)
during 3.92 d live time is $(8.3\pm0.4)$ events, rather low compared with the 96 measured events (see 
Tables \ref{tab2} and \ref{tab3}).

\begin{figure*}[!ht]
\begin{center}
\vspace{-0.6cm}
\resizebox{0.45\textwidth}{!}{\includegraphics{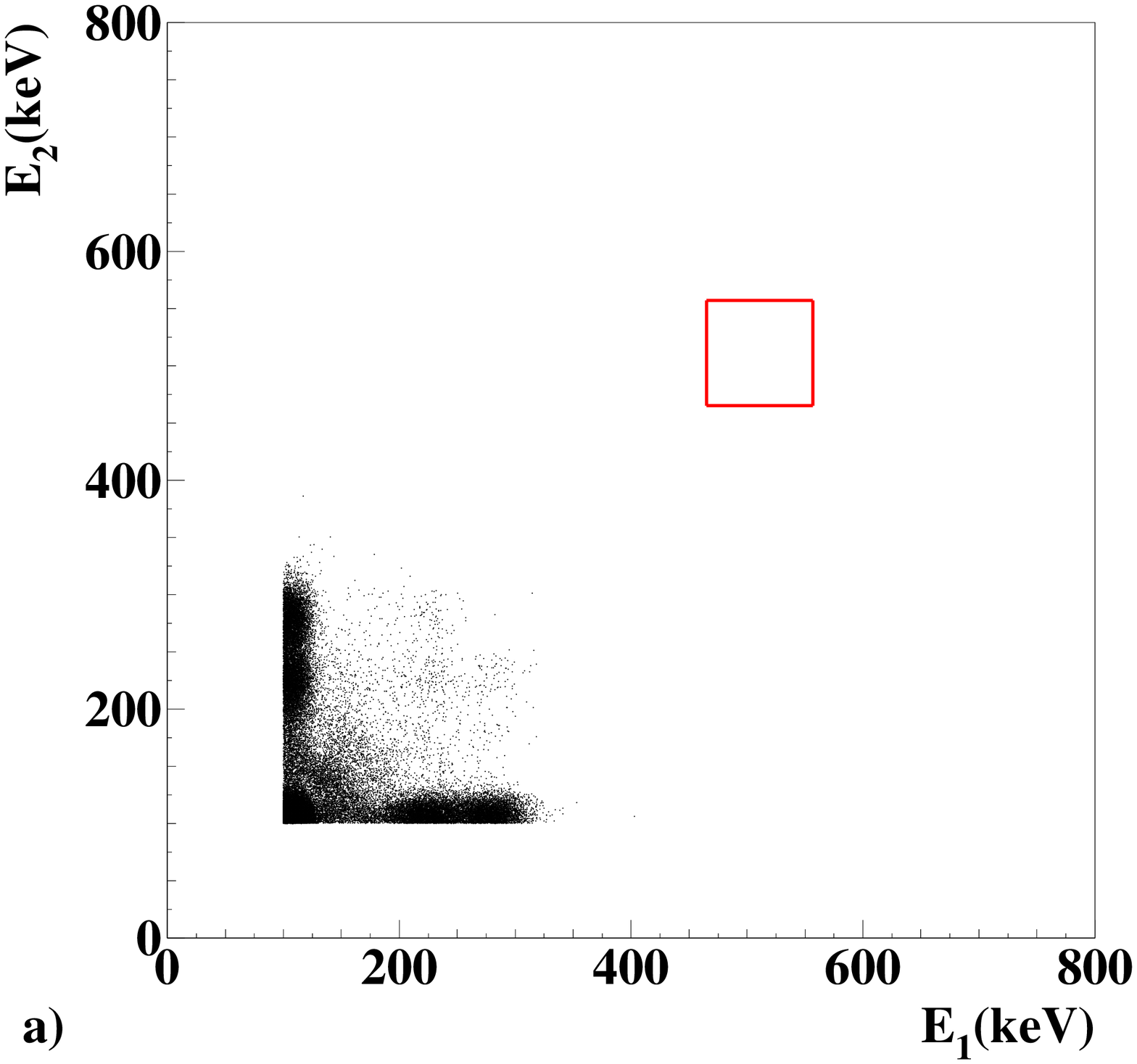}}
\resizebox{0.45\textwidth}{!}{\includegraphics{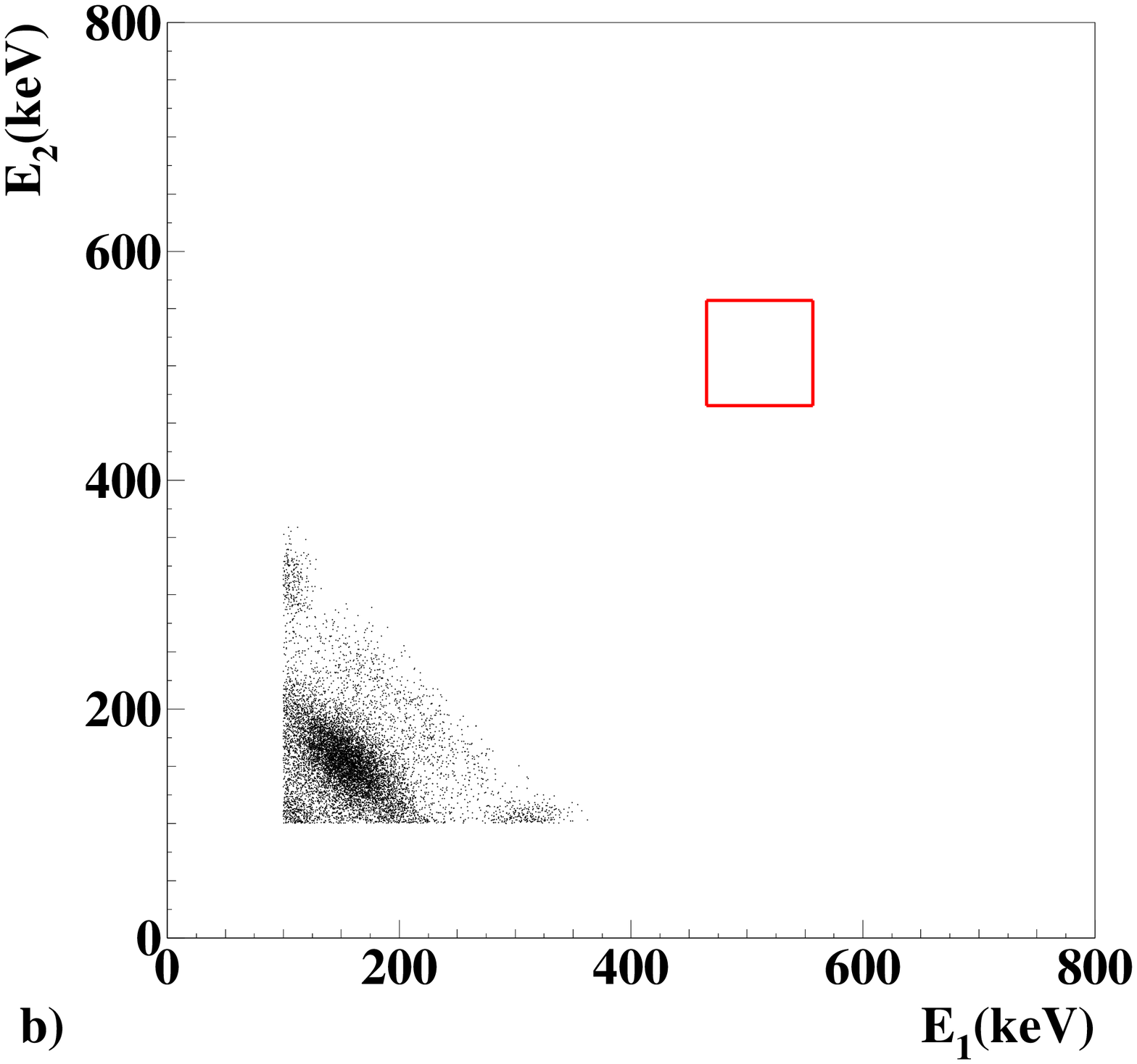}}
\end{center}
\vspace{-0.6cm}
\caption{($Color$ $online$) Model scatter plots of the $^{239}$Np (daughter of $^{243}$Am) decay (a) and of the $^{233}$Pa 
(in the $^{241}$Am radioactive chain) decay (b), obtained by the simulation. 
The red boxes define the region with $E_1$ and $E_2$ 
in the energy window (465--557) keV ($\pm 2 \sigma$ around 511 keV). The numbers of plotted events are arbitrary.}
\label{fig_oth}
\end{figure*}

As concerns the $^{243}$Am contribution, it $\alpha$ decays into $^{239}$Np, which $\beta$ decays with half-life 2.36 d
in the long living isotope $^{239}$Pu. The model scatter plots of the $^{239}$Np and of the $^{233}$Pa 
decays are depicted in Fig. \ref{fig_oth}. The former can explain the structures at $\simeq 110$ keV in coincidence with
$\simeq 220$ and $\simeq 280$ keV, present in the experimental data shown in Fig. \ref{fig1}a).
From the decay schemes of these two isotopes, one can see that no coincidence is expected in the region of 
interest for the present study.

\begin{figure}[!ht]
\begin{center}
\vspace{-0.6cm}
\resizebox{0.6\textwidth}{!}{\includegraphics{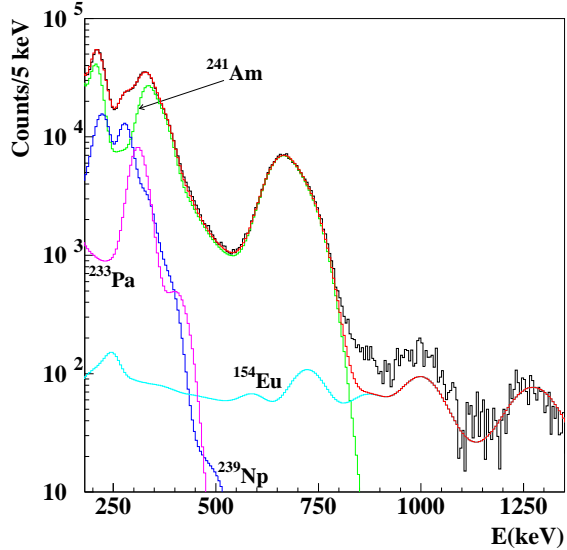}}
\end{center}
\vspace{-0.6cm}
\caption{($Color$ $online$) Experimental cumulative energy spectrum (black histogram) with multiplicity $\ge 1$, measured during 
3.92 d live time,
after the subtraction of the contribution of the set-up derived from the 24.6 d measurements without the sources.
The simulated contributions of $^{241}$Am (green histogram), $^{239}$Np (blue histogram), and $^{233}$Pa (purple histogram) are reported
normalizing their numbers of decays to their best fit values. The simulated contribution of $^{154}$Eu (light blue histogram) is
depicted for $1.70 \times 10^4$ $^{154}$Eu decays, see text. The red histogram is the sum of the four contributions.
A slight excess remains around 900 keV; see text.}
\label{fig_single}
\end{figure}

In order to estimate the amount of the aforementioned nuclides,
Fig. \ref{fig_single} shows the cumulative experimental energy spectrum (black histogram) with multiplicity $\ge 1$ measured during 3.92 d live time,
after the subtraction of the contribution of the set-up derived from the measurements without the $^{241}$Am sources. This energy spectrum
is fitted in the energy interval (195--770) keV by the model given by the sum of the mentioned components: 
$^{241}$Am, $^{239}$Np, $^{233}$Pa, and $^{154}$Eu.
The latter contribution is fixed to $1.70 \times 10^4$ $^{154}$Eu decays, as previously estimated.
For completeness we note that the contribution of the 511 keV events, observed in the coincidence scatter plot of 
Fig. \ref{fig1}a), is negligible and is not included in the fit.
The result of the fit is shown in Fig. \ref{fig_single} together with the four mentioned components.
A slight excess remains around 900 keV; it may be ascribed for example to 
a possible trace contamination (of order of 3.4 $\times$ 10$^{-2}$ ppt) of 
$^{234m}$Pa belonging to the $^{238}$U chain.

The best fit values for the total number of decays of $^{241}$Am, $^{239}$Np, and $^{233}$Pa are
$(4.717\pm0.011) \times 10^{10}$, 
$(1.228\pm0.007) \times 10^{6}$, and
$(2.81\pm0.04) \times 10^{5}$, respectively;
thus the average activity of $^{241}$Am in each source is $(34.88\pm0.08)$ kBq, compatible with the nominal 
activities of the sources\footnote{A small discrepancy ($\sim 5$\%) between measured and nominal values of the activity can be 
explained taking into account the error given by the producer (3.5\%) and the uncertainty on the estimated live time (few percent).
Anyway, the results in the following are largely independent on the live time (see later).} reported in Tables \ref{tab2} and \ref{tab3}.
Since the $^{243}$Am -- $^{239}$Np system is at equilibrium, 
the estimated average activities of $^{243}$Am and $^{239}$Np in each source are both equal to $(908\pm5)$ mBq;
this corresponds to $(443\pm3)$ ppm of $^{243}$Am contamination in the sources.
The $^{233}$Pa has instead an average activity in each source equal to $(208\pm3)$ mBq;
this value allows us to estimate the age of the sources, $t$, from their production. 
In fact, assuming that all the sources were produced at the same time and only $^{241}$Am isotopes ($N_0$ nuclides)
were present,
at the present time one can write: $N_{^{241}Am} = N_0 e^{-t/\tau}$ and $N_{^{237}Np} = N_0 (1-e^{-t/\tau})$,
where: i) $N_{^{241}Am}$ is the present number of $^{241}$Am nuclides, given by the best fit value of the 
average activity multiplied by $\tau$;
ii) $N_{^{237}Np}$ is the present number of $^{237}$Np daughters,
given by its activity (at equilibrium with $^{233}$Pa) multiplied by the mean-life of $^{237}$Np;
iii) $\tau=623.5$ yr is the mean-life of $^{241}$Am \cite{TOI98}.
By using these data, the age of the sources can be estimated as: $t=\tau \ln\left(1+\frac{N_{^{237}Np}}{N_{^{241}Am}}\right)=(18.1\pm0.3)$ yr, 
well in agreement with the specifications of the sources, also considering 
the production time of the sources material. 

\begin{figure}[!ht]
\begin{center}
\vspace{-0.6cm}
\resizebox{0.6\textwidth}{!}{\includegraphics{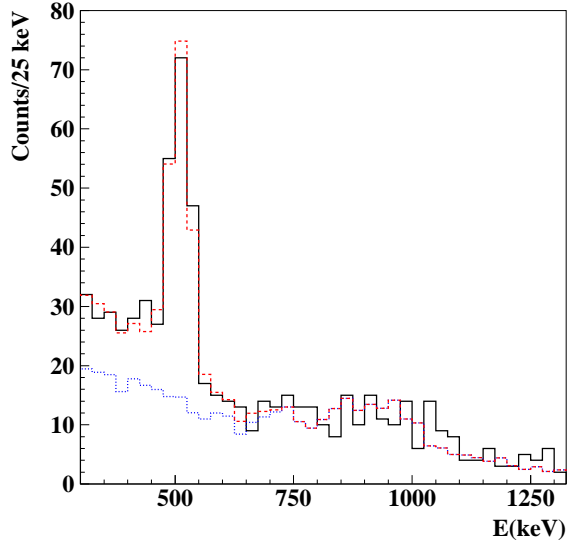}}
\end{center}
\vspace{-0.6cm}
\caption{($Color$ $online$)
Cumulative energy distributions measured by one detector of the pair when a coincident event with energy in the window 
(465--557) keV ($\pm 2 \sigma$ around 511 keV) is collected in the faced detector; both the detectors of the pair and 
all the used pairs contribute to the histograms.
$Solid$ $histogram$: experimental data in presence of the $^{241}$Am sources for the cumulative 3.92 d live time;
$dotted$ $histogram$ (blue online): model of the expected background obtained by summing the contributions
of the 24.6 d data without sources and of the $^{154}$Eu decays, both suitably normalized.
The presence of the 511 keV peak in the source data is well evident. The red dashed curve is the result
of the best fit (in the 300--750 keV region), that also includes a model of background and the contribution
of the positrons annihilations inside the sources.
The contribution from $^{241}$Am decay is negligible in this energy window. See text.}
\label{fig_fit}
\end{figure}

The excess of double coincidence events with energy around 511 keV in faced pairs of detectors seen in Fig. \ref{fig1}a)
can be analyzed to quantify the effect searched for. For this purpose, 
Fig. \ref{fig_fit} shows the cumulative energy distribution (solid histogram) measured by one detector of the pair when a 
coincident event with energy in the window (465--557) keV ($\pm 2 \sigma$ around 511 keV) is collected in the faced detector; 
obviously, both the detectors of the pair and all the used pairs contribute to the histogram.
As mentioned, the cumulative live time in presence of $^{241}$Am sources is 3.92 d.
In the same plot the dotted histogram represents the expected background obtained by summing the
contributions of the set-up (derived applying the same data 
selection to the data acquired in absence of the $^{241}$Am sources) and of the $^{154}$Eu decays.
The presence of a peak at 511 keV is clearly evident. Thus, a fit of the spectrum has been performed in the region 300--750 keV
by including\footnote{The $^{239}$Np and $^{233}$Pa decays give no contribution in this bidimensional energy window, as forementioned.}: 
i) the measured coincidences rate of the set-up in absence of the sources; ii) the determined coincidences rate of $^{154}$Eu decays;
iii) the known contribution from the $^{241}$Am decay, which is however negligible, as the decay scheme is known \cite{Bas06}; iv) a model of the background 
described by a straight line (this includes other possible contributions 
in the region of interest due to the sources, as e.g. possible $^{234m}$Pa); v) the simulated spectrum due to positrons 
annihilations within the sources. Thus the only free parameters are the 
straight line and the number of positrons annihilations within the sources.
The fit ($\chi^2$/d.o.f. = 3.7/14 = 0.26, where d.o.f. is the number of degrees of freedom) confirms the excess of
double coincidences from 511 keV $\gamma$ rays due to positrons annihilations.
The obtained number of positrons generated within the sources is: $N = (222 \pm 30)$.

Considering the $N$ positrons generated within the sources as related to IPP process, the $\lambda$ value can be derived 
according to the relation:
\begin{equation}
\lambda = \frac{A_{e^+e^-}}{A_{\alpha}} = \frac{N}{N_{\alpha}},
\label{lambda}
\end{equation}
where $N_{\alpha}$ is the total number of $^{241}$Am $\alpha$ decays during the running periods, whose best fit value
is $N_{\alpha} = (4.717\pm0.011) \times 10^{10}$ as previously estimated; it is worth noting that the $\lambda$ value 
is independent from the precise evaluation of the live time of the data taking.
Hence, the best estimate of $\lambda$ is $(4.70\pm0.63) \times 10^{-9}$, 
substantially compatible with the previous observations
and with the theoretical expectations (see Table \ref{tab1}). The coincidence excess in the energy region of interest 
is at 7.4 $\sigma$ C.L.

With the obtained estimate $N = (222 \pm 30)$ counts, we could derive also the cautious limit for the IPP process:
$\lambda < 5.5 \times 10^{-9}$ (90\% C.L.).

\vspace{0.3cm}

In the following, we have performed an investigation on the possible processes, other than $e^+e^-$ pair production in $^{241}$Am $\alpha$ decay, which 
-- totally or partially -- might account for the observed coincidence events.

\section{Investigation on possible coincidence background from the sources}

Let us summarize some relevant arguments:

\begin{itemize}

\item Any possible contribution to double coincidences from the used detectors (as e.g. from high energy $\gamma$'s interactions)
has been taken into account in the adopted procedure (see above). 

\item In the decay chain $^{241}$Am $\rightarrow ... \rightarrow ^{209}$Bi, $\beta^+$ emitters are absent \cite{TOI98}.

\item In the decay chain $^{241}$Am $\rightarrow ... \rightarrow ^{209}$Bi, the contribution of 
high energy $\gamma$'s (with energies larger than 1.022 MeV, that is able to produce $e^+e^-$ pairs) is negligible.
In particular, from the best fit values of the $^{241}$Am and $^{233}$Pa activities given above, one can obtain 
for the daughters' 
activities: 208 mBq for $^{237}$Np, 8.1 $\mu$Bq for $^{233}$U and 4.6 nBq for $^{229}$Th and its daughters. 
Taking into account these activities and safely assuming that each $\gamma$ is converted in a $e^+e^-$ pair, 
the contribution to $\lambda$ from high energy $\gamma$'s in the $^{241}$Am decay chain can be estimated as 
in the following:
i) 1110 keV $\gamma$ in $^{233}$U decay \cite{u233} can provide $\lambda < 2 \times 10^{-18}$;
ii) $\gamma$ lines at 1046, 1100, 1119 and 1328 keV in $^{213}$Bi decay \cite{bi213} 
can provide $\lambda < 2 \times 10^{-17}$, 
$\lambda < 3 \times 10^{-16}$, $\lambda < 7 \times 10^{-17}$ and 
$\lambda < 5 \times 10^{-19}$, respectively;
iii) $\gamma$ lines at 1239 and 1567 keV in $^{209}$Tl decay \cite{tl209} can provide $\lambda < 1 \times 10^{-17}$ and
$\lambda < 3 \times 10^{-15}$, respectively.

\item  No $\alpha$ decay of $^{241}$Am to an excited level of $^{237}$Np with energy larger than 0.8 MeV
has been observed to our knowledge and in the decay scheme of $^{241}$Am \cite{Bas06}\footnote{Upper limit on intensity 
of such decay was obtained in Ref. \cite{Sta86} as $< 10^{-9}$.}.

\end{itemize}

Thus, no direct contribution from the $^{241}$Am decay scheme \cite{Bas06} to double coincidences in the energy windows of interest 
is expected.
 
Remaining possible side processes are discussed in the following.

\subsection{Possible $\alpha-$induced reactions}

We recall that the used sources are made of a thin layer of active AmO$_2$ ($\phi \sim$ 3 mm) 
heat sealed by two thin mylar foils, and have been stored underground for $\gsim$ 10 yr. 

Thus, the elements that can give rise to ($\alpha$,n) or to ($\alpha$,p) reactions are the stable or long-lived isotopes of C, O (that are present in 
the source materials) and, if $\alpha$ particles may survive to the mylar foils, also N (gas surrounding the sources when they are inserted in the set-up) 
and Cu (housing of the NaI(Tl) detectors).

Considering the energy of the $\alpha$ particles emitted by the $^{241}$Am, the possible ($\alpha$,n) 
reactions are: $^{13}$C($\alpha$,n)$^{16}$O, $^{17}$O($\alpha$,n)$^{20}$Ne, $^{18}$O($\alpha$,n)$^{21}$Ne; none of the produced isotopes 
is a $\beta^{+}$ emitter. On the other hand, possible ($\alpha$,n) reactions on Ne produced by the O isotopes ($\alpha$,n) reactions:
$^{20}$Ne($\alpha$,n)$^{23}$Mg, can give $\beta^+$ emission but the energy threshold is 8.7 MeV, that is higher than the energy of the 
$\alpha$ particles emitted by $^{241}$Am.

Moreover, considering the energy threshold, the only possible ($\alpha$,p) reactions are:
$^{14}$N($\alpha$,p)$^{17}$O and $^{15}$N($\alpha$,p)$^{18}$O, both 
giving stable isotopes of O, and $^{63}$Cu($\alpha$,p)$^{66}$Zn and $^{65}$Cu($\alpha$,p)$^{68}$Zn
both giving stable isotopes of Zn.

A further source of background could be due to $e^+e^-$ pairs produced by photons from
($\alpha$,n$\gamma$) and ($\alpha$,$\gamma$) reactions
on the stable nuclei of C and O;
considering the measured cross sections \cite{spes,janis}, 
a cautious upper limit on this contribution can be derived: 
$\lambda_{e^+e^-}^{(\alpha,n\gamma),(\alpha,\gamma)} \ll 10^{-11}$.

In conclusion, possible significant contribution of $\beta^+$ emitters and of high energy $\gamma$'s 
from $\alpha$ interactions on known materials can be excluded; a similar study 
was performed in Ref. \cite{Lju73} where such possibilities were also excluded.

\subsection{$^{241}$Am fission fragments and possible cluster decays}

The spontaneous fission of $^{241}$Am has been measured in several works (see Ref. \cite{mar05} and reference therein) and the recommended lifetime 
is: $\tau_f = (1.7 \pm 0.4) \times 10^{14}$ yr.
Considering that the spontaneous fission activity is given by: $N_{^{241}Am}/\tau_{f}$, and that not all the fission fragments 
give rise to $e^+e^-$ emission, an upper limit:
$\lambda_{e^+e^-}^{fission} \ll 3.6 \times 10^{-12}$ can be derived. Thus, the spontaneous fission can be excluded
as source of the observed coincidence excess.

Another possibility may be the induced fission. 
From experimental studies \cite{shi99} the more probable fragments' masses in case of $^{241}$Am fission
induced by thermal neutrons (the one with higher cross section) are about 135 and 105.
Various nuclei with an excess of neutrons can be created with a cumulative yield up to 7\%, high $Q_{\beta}$ values (up to 8 MeV) and small 
$T_{1/2}$'s \cite{tnfis}. In their decay, deexcitations $\gamma$'s with high energies can be converted into $e^+e^-$ pairs and 
imitate the observed effect.
Thus, let us estimate the probability to have a fission from neutrons in the experimental conditions. 
The environmental flux of thermal neutrons inside the used multi-ton  multi-component shield\footnote{The flux of thermal neutrons outside the shield at LNGS is: 
$\sim 10^{-6}$ cm$^{-2}$s$^{-1}$ \cite{Bel89}, and the adopted neutron shield is made of $\sim$ 1 m concrete from the same Gran Sasso rock, 10/40 cm paraffin/polyethylene, 
1.5 mm Cd. In 
addition, there are 10 cm of OFHC Cu and 15 cm of Pb \cite{perflibra}.} has been measured to be: $< 1.2 \times 10^{-7}$ cm$^{-2}$s$^{-1}$ (90\% C.L.) \cite{perflibra}; 
the maximum cross-section for 
neutrons is $\sim$40 barn \cite{cal11,spes}, and the $^{241}$Am nuclei in a source are about $6.9 \times 10^{14}$. From these quantities a fission rate from environmental 
thermal neutrons surviving the shield can be derived to be: $< 10^{-14}$ s$^{-1}$; on the other hand, neutrons produced in this possible fission process
have a very low probability to induce further 
fissions 
($P\approx 10^{-7}$). Thus this process is too inefficient to account for the observations.

In addition to the environmental flux, neutrons can also be produced in the ($\alpha$,n) reaction on O isotopes; 
in particular, $^{17}$O and $^{18}$O which are in the mylar and in the AmO$_2$ with a total 
concentration of 
$3.5 \times 10^{15}$ cm$^{-2}$. However, considering a mean source activity of 35 kBq and a mean cross section for the process of 0.2 barn for $^{17}$O (natural abundance $\delta=0.038$\%) 
and 0.5 barn for $^{18}$O ($\delta=0.2$\%) 
\cite{spes}, a neutron production rate: $\sim 1.3 \times 10^{-7}$ s$^{-1}$, is derived with a neutron mean energy of order of MeV. The expected fission rate from these neutrons
($\sigma \sim 2$ barn \cite{spes}) is of order of $10^{-16}$ s$^{-1}$. Thus, in conclusion also this fission process is too inefficient to explain the observed excess.

Finally, another possible process is the clusters decay of $^{241}$Am.
The emission of $^{34}$Si is considered as the most probable one in the cluster decay of the $^{241}$Am: $^{241}$Am  $\rightarrow$ 
$^{34}$Si + $^{207}$Tl, with theoretical half lives: $3.2 \times 10^{17}$ yr \cite{tav07} and $5.6 \times 10^{18}$ yr \cite{bha08}.
Both the created fragments are unstable. In the decay of $^{207}$Tl, only the ground state and few excited levels of $^{207}$Pb with $E_{exc} < 1022$ 
keV are populated; their deexcitation cannot produce $e^+e^-$ pairs. 
However, in the decay chain $^{34}$Si ($T_{1/2} = 2.77$ s, $Q_{\beta^-} = 4601$ keV) $\rightarrow$ $^{34}$P ($T_{1/2} = 12.43$ s, $Q_{\beta^-} = 5374$ 
keV) $\rightarrow$ $^{34}$S \cite{TOI98} much more energy is released, and the deexcitation of high energy levels in $^{34}$P and $^{34}$S leads 
to emission of $e^+e^-$ pairs.
The best current experimental limit on the decay: $^{241}$Am  $\rightarrow$ $^{34}$Si + $^{207}$Tl, is $T_{1/2} > 5.8 \times 10^{17}$ yr 
\cite{moo87}; the limit for other clusters emission is $ > 8.6 \times 10^{16}$ yr \cite{mar05}.
In conclusion, the possible contribution of $^{241}$Am cluster decay is negligible ($\lambda < 5 \times 10^{-15}$) and cannot account for the 
observed effect.

\section{Conclusions}

The experimental data show an excess of double coincidences in the 511--511 keV energy region which are not 
explained by known analysed background processes as discussed above. Moreover, 
any sizeable contamination of the sources themselves by isotopes $\beta^+$ emitters is not present to our knowledge.

Thus, this measured excess gives a a relative activity
$\lambda = (4.70 \pm 0.63) \times 10^{-9}$ for the Internal Pair Production (IPP)
with respect to alpha decay of $^{241}$Am; this value is of the same order of magnitude as the previous 
determinations \cite{Lju73,Sta86,Asa90} obtained by using different 
setups, sources (with different features and producers) and experimental approaches.
In a conservative approach the upper limit $\lambda < 5.5 \times 10^{-9}$ (90\% C.L.) can be derived.

For completeness, we mention that, if one considers the exotic frame hypothesized in Ref. \cite{Ino90}, one 
would get $\lambda$ value or $\lambda$ limit of the same order of magnitude as those given above.

Further investigations are foreseen in the future with different dedicated sources and set-up.

Finally, it is worth noting that this is the first result on IPP obtained in an underground
experiment (in the previous ones the sea-level cosmic-rays background was estimated and subtracted), and
that the $\lambda$ value obtained in the present work is independent on the
live-time estimate (which is difficult to estimate with very high accuracy in similar experiments because of
the relatively high intensities of the $^{241}$Am sources).

\section{Appendix}

In order to estimate a possible presence of a coincidence excess in the region of interest in runs with presence of $^{241}$Am sources with respect to runs without 
sources, the following procedure has been adopted.

Let us consider that $n_1$ is the number of coincidence events measured in the region of interest in presence of the source during a live time $T_1$. 
The conditional probability to measure $n_1$ coincidence events in presence of a signal $S=s T_1$ ($s$ is the signal rate, see text) 
and of a background rate $b$ is given by:
\begin{equation}
P(n_1 \mid S+bT_1) = \frac{e^{-(S+bT_1)}(S+bT_1)^{n_1}}{n_1!}.
\end{equation}

The $b$ parameter can be estimated by performing a run in absence of sources, where $n_2$ coincidence events are collected
in a live time $T_2$. In fact, the conditional probability to obtain $n_2$ given a background rate $b$ is:
\begin{equation}
P(n_2 \mid bT_2 ) = \frac{e^{-(bT_2)}(bT_2)^{n_2}}{n_2!}.
\end{equation}
To determine the probability distribution function  ($p.d.f.$) of $S$, ${\mathcal L}(S)$, the procedure of 
Ref. \cite{Hel83} can be followed. If the {\it prior} probability of $b$ is a uniform function in $(0,\mathcal 1)$, 
the {\it posterior} $p.d.f.$ -- given $n_2$ -- can be written by using the Bayes' theorem:
\begin{equation}
f(bT_2 \mid n_2) = A' P(n_2 \mid bT_2), 
\end{equation}
while the {\it posterior} $p.d.f.$ for $S$ -- given $n_1$ and the $b$ parameter -- is:
\begin{equation}
f(S \mid n_1,bT_1) = A'' P(n_1 \mid S+bT_1)
\end{equation}
where $A'$ and $A''$ are normalization constants. 
Since $b$ is a nuisance parameter, one can marginalize with respect to it obtaining:
\begin{eqnarray}
{\mathcal{L}}(S)=\int_0^{\mathcal1}f(S \mid n_1 , bT_1)f(bT_2 \mid 
n_2)d(bT_2) = \nonumber \\
= A \int^{\mathcal1}_0 e^{-(S+bT_1)} (S + bT_1)^{n_1} 
e^{-bT_2} (bT_2)^{n_{2}} d(bT_2)
\end{eqnarray}
where $A$ is a normalization constant such that 
\begin{equation}
\int_0^{\mathcal1} {\mathcal L}(S) dS = 1.
\end{equation}

These integrals can be analytically solved and the obtained results are:
\begin{equation}
{\mathcal L} (S) = A e^{-S} \sum_{k=0}^{n_1} {n_1 \choose k} 
S^{n_1-k} \eta^{k} \frac{(k+n_2)!}{(\eta +1)^{n_2+k+1}}.
\end{equation}
where $\eta = T_1 / T_2$. The constant $A$ is given by:
\begin{equation}
A = \frac{(1+\eta)^{n_2 +1}}{n_1!n_2!\displaystyle \sum_{k=0}^{n_1} 
{n_2+k \choose n_2} \left( \frac{\eta}{\eta +1} \right) 
^k}.
\end{equation}

Finally, according to the Wilks' theorem,
the quantity $-2\ln({\mathcal L}(S))$ follows a $\chi ^2$ distribution \cite{pdg}; thus to obtain the best estimate for $S$ and 
its interval of confidence, the same procedure 
used for the $\chi ^2$ distribution can be followed.

\end{document}